\documentclass{JHEP}
\usepackage{epsfig}

\preprint{
IMSc/2002/02/04\\}
\title{On the UV renormalizability of noncommutative field theories}
\author{Swarnendu Sarkar \\
The Institute of Mathematical Sciences,\\
CIT Campus,
Chennai - 600 113,  INDIA \\
\email{swarnen@imsc.ernet.in }}

\abstract {
UV/IR mixing is one of the most important features of noncommutative field
theories. As a consequence of this coupling of the UV and IR sectors, the
configuration of fields at the zero
momentum limit in these theories is a very singular configuration. 
We show that the renormalization conditions set at a particular
momentum configuration with a fixed number of zero momenta, renormalizes
the Green's functions for any general momenta only when this configuration
has same set of zero momenta. Therefore only when renormalization
conditions
are set at a point where all the external momenta are nonzero, the quantum
theory is renormalizable for all values of nonzero momentum.
This arises as a result of different scaling behaviors of Green's
functions with respect to the UV cutoff ($\Lambda$) for configurations
containing
different set of zero momenta.
We study this in the noncommutative $\phi^4$ theory and analyse similar
results for the Gross-Neveu model at one loop level. 
We next show this general feature using Wilsonian RG of Polchinski 
in the globally $O(N)$ symmetric scalar theory and prove the 
renormalizability of the theory to all orders with an infrared cutoff.
In the context of spontaneous symmetry breaking (SSB) in noncommutative
scalar theory, it is essential to note the different scaling behaviors of
Green's functions with respect to $\Lambda$ for different set of zero
momenta configurations. We show that in the broken phase of the theory 
the Ward identities are satisfied to all orders only when one keeps an 
infrared regulator by shifting to a nonconstant vacuum. }

\newcommand{\bel}{\begin{equation}\label}
\newcommand{\f}{\frac}
\newcommand{\non}{\nonumber \\}
\newcommand {\beq}{\begin{equation}}
\newcommand {\eeq}{\end{equation}}
\newcommand {\beqa}{\begin{eqnarray}}
\newcommand {\beqal}{\begin{eqnarray}\label}
\newcommand {\eeqa}{\end{eqnarray}}
\newcommand {\bc}{\begin{center}}
\newcommand {\ec}{\end{center}}

\def\vs5{\vspace*{5mm}}
\def\vs1{\vspace*{1cm}}
\def\vs2{\vspace*{2cm}}
\def\hs5{\vspace*{5mm}}
\def\hs1{\hspace*{1cm}}

\begin{document}

\section{Introduction}

Noncommutative spacetimes and field theories defined on them have been
studied extensively for the past few years mainly motivated from
string theory \cite{connes,bal,witten}. 
Apart from this fact that these theories arise as low energy
limits of string theory in a constant $B_{\mu\nu}$ background, their study
as field theories in their own right is quite facinating . For reviews on
the subject see \cite{ncft1,ncft2,ncft3}. 
Various interesting aspects of these theories have been studied rigorously
recently \cite{filk}-\cite{ruiz}.
The most important of these being the intriguing mixing of UV
and IR
divergences which is a direct consequence of the noncommutavity of the
background spacetime. Perturbative studies of these field theories carried
out extensively revealed various nontrivial aspects arising from this
transmutation of UV into IR divergences . One of the most important being
the alteration of the conventional Wilsonian picture of renormalization
group flows in the very low momentum domain.
The UV renormalizability of these theories has been
argued \cite{min} and in some cases expiicitly shown upto two loops
\cite{twoloop}. The
$\lambda\phi^4$ theory has been shown to be renormalizable as long as the
external momenta, $p$ for the $n$-point functions are such that
$\Lambda^2pop
>1$.\cite{wilson}.
However given that we are only interested in the continuum limit,
$\Lambda \rightarrow \infty$, the inequality is not satisfied only
when $p$ is restricted to the zero momentum value.

The  possibility that these theories would ultimately be defined
with an infrared cutoff still exists. It was shown \cite{phase} that phase
transitions
if possible can only occur at a finite momentum leading to a 
nonhomogeneous phase. 

Normally in commutative field
theories the renormalization conditions required to absorb the infinities
in the Green's functions of a particular configuration of fields, 
with momenta around some scale, leads to the infinities being absorbed
from the Green's functions at all scales i.e. the functional
dependence of the divergent $n$-point functions on the UV cutoff
$\Lambda_0$ is same at all values of external momenta. 
On the other hand, in noncommutative theories,
due to coupling of the UV and IR sectors, the zero momentum limit of
a particular configuration of fields is singular. We show that
Greens's functions scale with different coefficients of $\Lambda$ for
configurations which differ by the number of zero external momenta. If the
renormalization conditions are set at a point where all the external
momenta are nonzero, then the bare couplings defined through this would
have a different $\Lambda_0$ dependence from the case when the
renormalization point consists of a number of zero momenta.
This shows that with the former renormalization condition, the theory for
all values of external momenta, $p \ne 0$ is renormalizable, while the
latter leads to a nonrenormalizable theory. 
We discuss this issue in the
noncommutative $\phi^4$ theory as well as in  the
Gross-Neveu model as shown in \cite{gn}. 

We next study the renormalizability of the globally $O(N)$
symmetric noncommutative scalar theory in its symmetric and its broken
phases to all orders. First we review the same for the commutative case
in the renormalization group approach.\cite{sharat}\cite{pol}
The noncommutative theory is then proved to be renormalizable to all
orders with an infrared cutoff. Keeping in mind the observations stated in
the previous paragraph,
we seperate the sector with external momenta such that $\Lambda^2 pop
>1$ from that of  $\Lambda^2 pop < 1$. 
In the latter case, since in the continuum limit $p$ is only
restricted to zero, the full quantum theory for nonzero $p$ will thus be
defined with an Infrared cutoff. Because of the presence of the infrared
cutoff for the external momenta, we shall also formally introduce an IR
cutoff for the internal loop momenta $(\Lambda_{IR})$. However we shall
see that the in the loop computations the IR cutoff for the internal
momenta is not necessary as there are no IR divergences in the loop
intergals in this RG approach. IR divergences appear in the continuum
limit as the zero momentum field configuration is approached from a
nonzero momentum field configuration. Ofcourse with an IR cutoff in the
external momenta, $p$ such that, $\Lambda^2 pop >1$, there are no IR
divergences in the theory. However IR divergences do appear in
perturbation theory. This is illustrated by an example of a 2-point
diagram computation. 

It may seem that $(\Lambda_{IR})$ may be needed, so that the
the cannonical scaling of the relevant and irrelevant operators are not
affected by the UV/IR transmutation. However this is not the case. This
point will be clarified in section 4.    

We then demonstrate
how the different scaling behaviors
of the Green's functions for configurations having different set of zero
momenta plays a crucial role when proving
the renormalizability of the spontaneously broken phase.
It was shown in \cite{my} that the renormalizability of the broken
symmetric phase to one loop 
could be proved by shifting to a phase where the vacuum was a nonconstant
background field. This nonconstant vacuum acts as an infrared regulator. 

This paper is organised as follows. In section 2 we study the
noncommutative $\phi^4$ theory and show that the UV cutoff ($\Lambda$)
dependence of the bare couplings are different depending on whether we set
the renormalization conditions at $p=0$ or $p\ne 0$. In section 3 we
analyse some more results
for the noncommutative Gross-Neveu model along the same line as section 2.
In section 4 we study
SSB in noncommutative scalar theory with global $O(N)$ symmetry in its
symmetric phase. We first review the one loop
results from \cite 
{ssb,my}. Next we prove the renormalizability of the
symmetric phase of the globally $O(N)$ symmetric noncommutative scalar
theory to all orders, after
reviewing the same for the commutative case.
For the broken phase we
demonstrate that, by going to a phase which is
translationally noninvariant i.e. by keeping the shift $v$ as a
nonconstant background field one is able to work with a infrared regulator
so that the problem of different scaling behaviors of Green's functions
for nonzero external momenta from that of the zero momentum case does not
arise. With this infrared regulator we prove the renormalizability of the
broken
phase of the theory to all orders. 
 We give our conclusions in section 5.

\section{UV/IR mixing and UV renormalizability}

In this section we show that the one loop Green's functions defines
two different renormalization conditions depending on whether we set
$p=0$ right in the begining or approach this limit with a nonsingular,
$p\ne 0$ configuration. We study
the noncommutative $\phi^4$ theory to show this. For an introduction to
noncommutative scalar theory see \cite{min}. 

The lagrangian for the theory is,

\beqa
{\cal L}_E=-[\f{1}{2}{(\partial_{\mu}\phi)}^2 + \f{1}{2}m^2\phi^2 +\f{\lambda}{4}\phi*\phi*\phi*\phi]
\eeqa

where,
\beq
\phi_1*\phi_2=e^{\f{i}{2}\partial_{\mu}^{y}\theta^{\mu\nu}\partial_{\nu}^{z}}
\phi_1(y)\phi_2(z) \big |_{
y=z=x}
\eeq

and $\theta$ is an antisymmetric matrix.
The propagator for the theory, is same as that of the commutative theory. Only the interaction term has a nontrivial momentum dependence. The inteaction vertex is given by,

\beqal{v}
-2\lambda V({\bf p})=-2\lambda[cos(p_1\wedge p_2)cos(p_3\wedge p_4) &+&
cos(p_1\wedge p_3)cos(p_2\wedge p_4) \non &+& cos(p_1\wedge p_4)cos(p_2\wedge p_3)]
\eeqa

The only divergent functions for the theory are the two point and the four 
point functions. The diagrams corresponding to the one loop contributions 
to these functions are shown in figures 1 and 2.

\begin{figure}[htbp]
\begin{center}
\epsfig{file=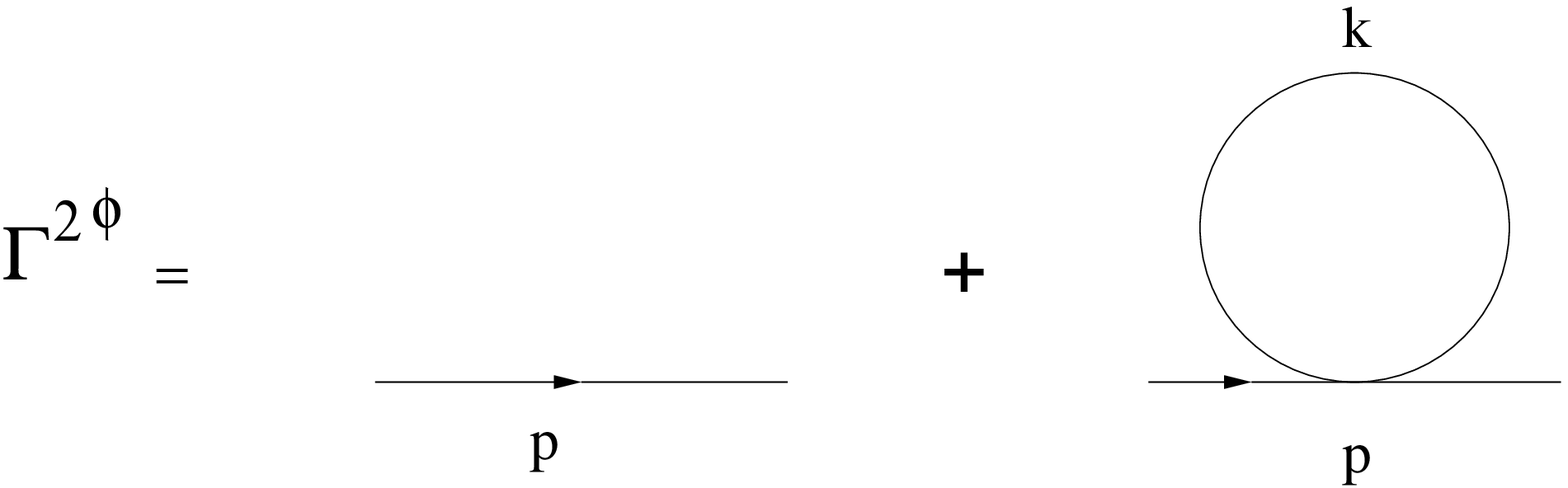, width=8 cm, angle=0}
\end{center}
\caption{}
\end{figure}

\beqa
\Gamma^2=-[p^2+m^2] -\lambda\int\f{d^4k}{(2\pi)^4}\f{2+cos(p\wedge k)}{k^2 + m^2}
\eeqa

The $cos$ term inside the integral regulates the second part of
the integral and is finite for $p\ne 0$.

\beqa
\Gamma^2=-[p^2+m^2]
-\f{\lambda}{8\pi^2}[\Lambda^2-m^2ln(\f{\Lambda^2}{m^2}) +O(1)]\non
-\f{\lambda}{16\pi^2}[\Lambda^2_{eff}(p)-m^2ln(\f{\Lambda^2_{eff}(p)}{m^2}) 
+O(1)]
\eeqa

where,

\beqa
\Lambda^2_{eff}(p) = \f{1}{\f{1}{\Lambda^2}+pop}\non
pop = -\f{p^{\mu}\theta^2_{\mu\nu}p^{\nu}}{4}
\eeqa

$\Lambda$ is the UV cutoff. We shall call terms containing
$\Lambda^2_{eff}(p)$ as nonplanar terms.

\begin{figure}[h]
\begin{center}
\epsfig{file=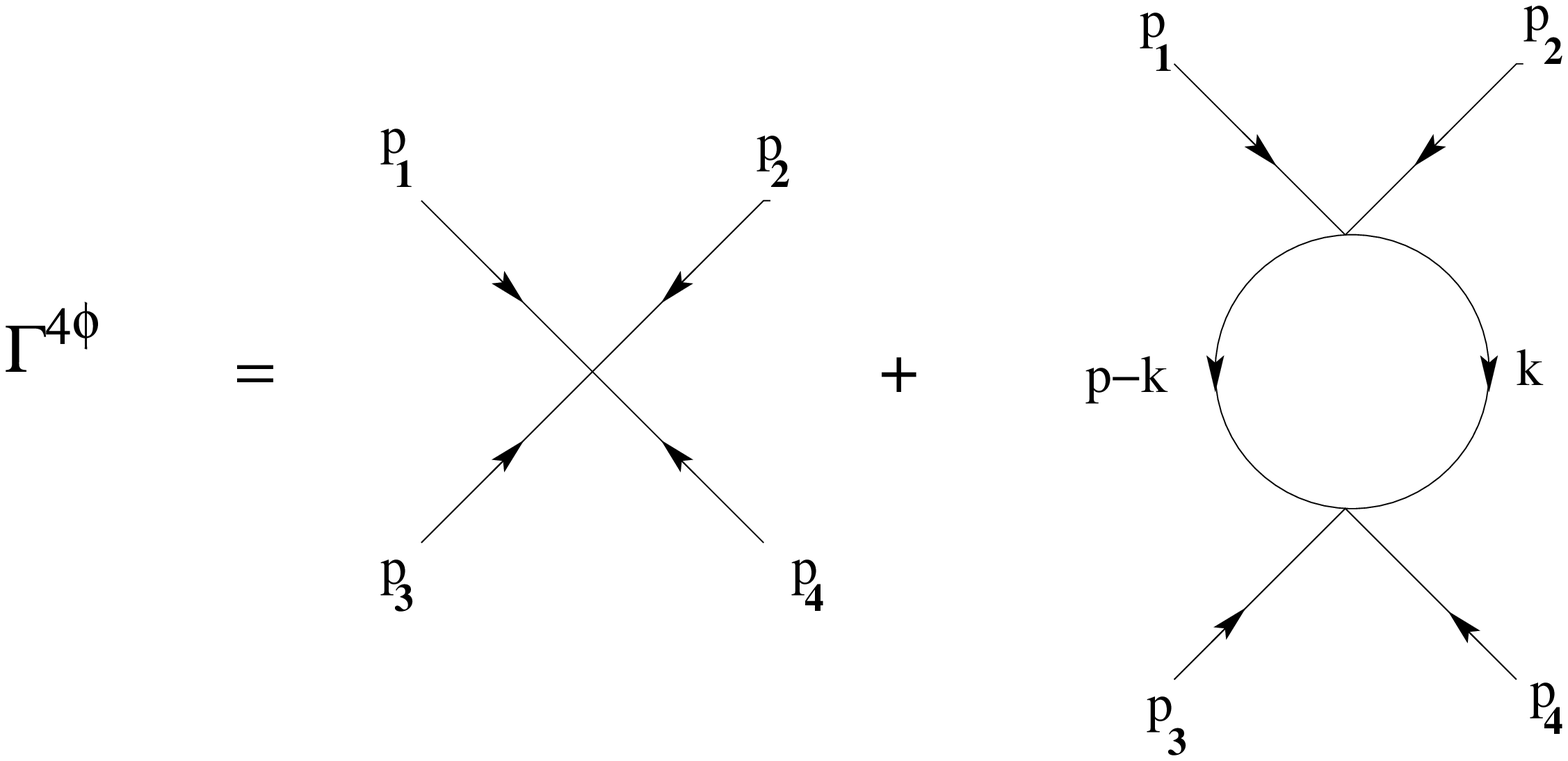, width= 8 cm,angle=0}
\begin{caption}
{  }
\end{caption}
\end{center}
\end{figure}

\beqal{gamma4}
\Gamma^4=-2\lambda V({\bf p}) +2\lambda^2\int \f{d^4k}{(2\pi)^4}\f
{F(p_1,p_2,p_3,p_4,p,k)}{(k^2+m^2)[(p-k)^2+m^2]} + \mbox{ s and u 
channels}
\eeqa
\noindent
where $F(p_1,p_2,p_3,p_4,p,k)$ is a function of terms containing {\it cos}
of the external momenta.

\beqa
\Gamma^4=-2\lambda V({\bf p}) + 4\lambda^2V({\bf p})\int
\f{d^4k}{(2\pi)^4}\f {1}{(k^2+m^2)[(p-k)^2+m^2]}+ NP
\eeqa

\noindent
where $NP$ are the Nonplanar terms. These terms would give rise to IR 
divergences as the external momenta goes to zero. 
The UV divergent piece for this four point amplitude is given by,

\beqa
\Gamma^4 \sim \f{\lambda^2}{4\pi^2}V({\bf p})ln(\f{\Lambda^2}{m^2})
\eeqa

The two and the four point contributions to the effective action to one
loop is,

\beqa
-
S_{eff}&=&\f{1}{2!}\int dp\phi(p)\phi(-p)[p^2 +m^2
+\f{\lambda}{8\pi^2}\Lambda^2-\f{\lambda}{8\pi^2}m^2ln(\f{\Lambda^2}{m^2})  +
NP ]\non
&+&\f{1}{4!}\int dp_1 dp_2 dp_3 dp_4[P + NP]
\phi (p_1)\phi (p_2)\phi (p_3)\phi (p_4)\delta (\sum p_i)
\eeqa
\noindent
where, $P$ is the planar term from the four point amplitude given by,

\beqa
P=V({\bf p})\{2\lambda -\f{\lambda^2}
{4\pi^2}ln(\f{\Lambda^2}{m^2})\}
\eeqa

\noindent
The renormalised parameters may now be defined as,

\beqal{rcs1}
m^2_{R} &=& m^2
+\f{\lambda}{8\pi^2}[\Lambda^2-m^2ln(\f{\Lambda^2}{m^2})]\non
\lambda_R&=&\lambda -\f{\lambda^2}{8\pi^2}ln(\f{\Lambda^2}{m^2})
\eeqa

As the zero momentum limit is approached, the $NP$ terms in
equation (\ref{gamma4}) 
give rise
to $IR$ divergences, however the renormalization conditions in
equation (\ref{rcs1}) lead to
a UV renormalizable quantum theory at one loop.

Now let us consider the case where the effective action at the zero
momentum field configuration is defined by \cite{coleman},

\beqal{veff}
V_{eff}=\sum_{n=1}^{\infty}\f{1}{n!}\Gamma^{n}(0,0,...,0)\phi^n
\eeqa

We use this to define $\lambda(\Lambda)$. In this case at one loop level
there are no nonplanar diagrams and the
potential to one loop is exactly equal to the commutative theory. The
external momenta are all put to zero before all loop calculations. To one
loop the effective potential is given by,

\beqa
V_{eff}&=&\f{1}{2}m^2\phi^2+\f{\lambda}{4}\phi^4
+\f{1}{2}\int^{\Lambda}\f{d^4k}{(2\pi)^4}ln(1+\f{3\lambda\phi^2}{k^2+m^2})\\
&=&\f{1}{2}m^2\phi^2+\f{\lambda}{4}\phi^4 +
\f{3\lambda\phi^2}{32\pi^2}[\Lambda^2 - m^2 ln(\f{\Lambda^2}{m^2})]
+ \f{9\lambda^2}{64\pi^2}\phi^4ln(\f{3\lambda\phi^2}{\Lambda^2})
\eeqa

\noindent
where in the final expression we have dropped terms which have
negative powers or are independent of $\Lambda$.

The renormalized quantities would now be defined by,
 
\beqa
\f{d^2V}{d\phi^2} \big |_{ \phi=0}&=&m^2_{R}\\
\f{d^4V}{d\phi^4} \big |_{ \phi=\phi_0}&=&6\lambda_{R}
\eeqa

These lead to,

\beqal{rcs2}
m^2_{R}&=&m^2+\f{3\lambda}{16\pi^2}[\Lambda^2 - m^2
ln(\f{\Lambda^2}{m^2})]\non
\lambda_R&=&\lambda-\f{9\lambda^2}{16\pi^2}ln(\f{\Lambda^2}
{3\phi_0^{2}})
\eeqa

From equations (\ref{rcs1}) and (\ref{rcs2}) it is clear that the
renormalization conditions
defined by equation (\ref{rcs2}) would not lead to a renormalizable
noncommutative theory for
non zero external momenta. 
As noted before, the configuration of fields with $p = 0$ is
singular and the two different renormalization conditions
(\ref{rcs1}), (\ref{rcs2}) occur as a
consequence of setting  $p = 0$ right in the begining or of 
approaching this configuration as a limit $p\rightarrow 0$. This is a
generic feature of noncommutative theories. The origin of this is the
transmutation of UV divergences into the IR divergences. 

In terms of the renormalization group flows, equations
(\ref{rcs1}), (\ref{rcs2}) states that
at the one loop level the relevant (relevant plus marginal) coupling
$\lambda$ scales with respect
to $\Lambda$ with different coefficients. This  means that the
functional dependence of the bare coupling on the UV cutoff $\Lambda$ are
different in the two cases. We shall see this in section 4, where the relevant 
couplings would scale with different coefficients of $\Lambda$ depending
on whether or not
the external momenta are such that $\Lambda^2pop<<1$.

\section{The noncommutative Gross-Neveu model}

In this section we review some of the results of the
noncommutative Gross-Neveu model \cite{gn}
which are along the same line as those of the previous section.

The lagrangian for the noncommutative Gross-Neveu model is, 

\beqa
{\cal L}_{E}=-[\f{1}{2}{\bar \psi^i} \gamma^{\mu}\partial_{\mu} \psi^i
+\f{\lambda}{8N}{\bar
\psi^i}*\psi^i*{\bar \psi^j}*\psi^j]
\eeqa

Where $\psi$ is a $2$-component spinor and $\gamma^{\mu}$ are $2\times 2$
Dirac matrices.
To evaluate the large $N$ limit of the effective action it is helpful to
introduce an auxillary field $\sigma$, so that,

\beqa
{\cal L}_{E}=-[\f{1}{2}{\bar \psi^i}\gamma^{\mu}\partial_{\mu}
\psi^i-\f{8N}{\lambda}\sigma^2 -2\sigma*{\bar \psi^i}*\psi^i]
\eeqa

\begin{figure}[htbp]
\begin{center}
\epsfig{file=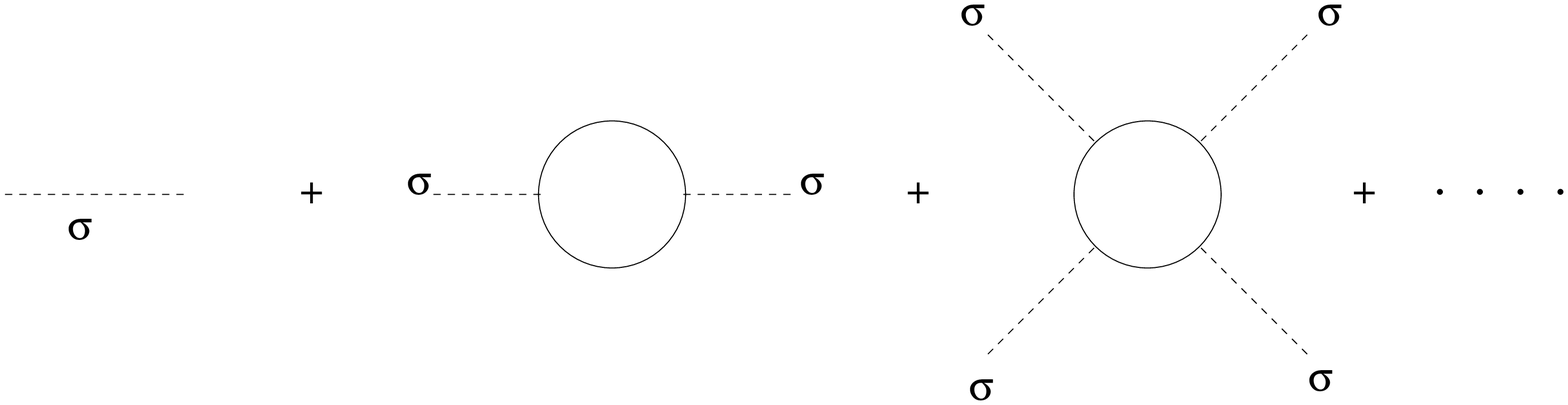, width=10 cm, angle=0}
\end{center}
\caption{}
\end{figure}

We now evaluate the effective action at the zero momentum field
configuration as defined by equation (\ref{veff}). 
The processes with only $\sigma$ field on the external legs would
contribute to the lowest order in $\f{1}{N}$ \cite{gross}.
These diagrams contributing to this lowest order are shown
in figure 3.

\beqa
V_{eff}&=&\f{8N}{\lambda}\sigma^2-N\sum_{n=1}^{\infty}\int\f{d^2k}{(2\pi)^2} 
\f{1}{2n}(\f{4\sigma^2}{k^2})^n\\
&=&\f{8N}{\lambda}\sigma^2
-\f{N\sigma^2}{2\pi}[ln(\f{\Lambda^2}{4\sigma^2})+ O(1)]
\eeqa

The renormalization condition,

\beqa
\f{\partial^2 V_{eff}}{\partial \sigma^2}{\big |}_{\sigma^2=\sigma^2_0}=1
\eeqa
defines the renormalized coupling $\lambda_R$ as,

\beqal{rcg1}
\f{1}{\lambda_R}=\f{1}{\lambda}-\f{1}{16\pi}ln(\f{\Lambda^2}{4\sigma^2_0})
\eeqa

 We now write down the effective action for nonzero external momenta in
the large $N$ limit. Only the two point function is divergent. To extract
the renormalization condition it is sufficient to  evaluate the
effective action with contributions upto the two point function only.

\beqa
S_{eff}=\f{1}{2}\int d^2p \sigma(p)\Gamma^{\sigma\sigma}\sigma(-p)
\eeqa

where,

\beqa
\Gamma^{\sigma\sigma}= \f{8N}{\lambda}-N\int
\f{d^2k}{(2\pi)^2}\f{k.(p+k)}{k^2(k+p)^2} + NP
\eeqa

$NP$ is the nonplanar term. The renormalized coupling will now be defined
as,

\beqal{rcg2}
\f{1}{\lambda_R}=\f{1}{\lambda}-\f{1}{32\pi}ln(\f{\Lambda^2}{p^2})
\eeqa

The renormalization condition, equation (\ref{rcg2}) gives a UV
renormalizable quantum theory
for any nonzero value of external momenta, similar to the results obtained
in the
previous section. If one defines the renormalized coupling as equation
(\ref{rcg1}), the
quantum theory at nonzero external momenta is nonrenormalizable. These two
renormalization conditions correspond to two different bare theories as 
argued at the end of section 2.

The ground state of the theory will be defined by a particular configuration
of fields. If it is defined with zero external momenta, we have noted that
this is a very singular field configuration and Green's functions in this
configuration has a different $\Lambda$ dependence from the nonzero
momenta field configurations. If the renormalization conditions are set at
zero momentum the theory is nonrenormalizable. 
In the next section we shall see how this
plays a crucial role in deciding the
renormalizability of the Broken Phase in a spontaneously broken global   
symmetric theory.

\section{SSB in Non-commutative scalar theory}

In the previous two sections we have seen that due to the singular
behavior of the IR limit in nncommutative field theories, the Green's
functions have different $\Lambda$
dependences  for nonzero and zero momentum field configurations. In this
section we show how this plays a crucial role when proving
renormalizability of the broken phase of a spontaneously broken
globally $O(N)$ symmetric theory. 
We first review our
one loop results \cite{my} and then in the later part of this section we 
shall prove following \cite{pol}, the renormalizability of the symmetric
phase
as long as the external momenta $p \ne 0$ such that
$\Lambda^2pop>1$. Formally with an IR cutoff in the external momenta, one
should also introduce an IR cutoff for the internal loop
momenta $(\Lambda_{IR})$. The presence of this IR cutoff in the loop
computations would be implicitly assumed. This point is elaborated in
section 4.2. 
The relevant couplings for the zero
momentum configuration of fields scale with different coefficients of the
UV cutoff from the $p \ne 0$ configurations. If the renormalization
conditions are set in these two momentum regimes, the bare couplings will
have different $\Lambda$ dependences.
At the end of the section we shall show the renormalizability of the
broken phase of the theory to all orders in terms of the
renormalization group.

\subsection{One loop analysis}

We consider a scalar field theory
with global $O(2)$ symmetry in its symmetric phase in 4 dimensions. 
We refer the reader to \cite{my} for the details of calculations.

\setcounter{footnote}{0}
The lagrangian for the theory is\footnote{Note that in the
noncommutative theory, there are various possible inequivalent orderings
of the fields for the quartic term. One of these terms has been chosen as 
an example. The proof of the renormalizability of the globally $O(N)$
symmetric theory, in its symmetric as well as in its broken phases, to
all orders, outlined in the latter part of this section remains unaltered 
for any such $O(N)$ symmetric quartic term.},

\beqal{lsy}
L_S &=& -[\f{1}{2}(\partial_{\mu}\phi^i)^2 - \f{1}{2}\mu^2(\phi^i)^2 +
\f{\lambda}{4}\phi^i*\phi^i*\phi^j*\phi^j] \non
i,j&=&1,2
\eeqa 

Here $\mu^2 > 0$, so that at the tree-level the theory undergoes SSB. The
global O(2) symmetry of the quantum theory implies that the green's
functions satisfy the following set of Ward identities in the symmetric
phase,

\beqal{wi1}
{\f{\delta^2\Gamma}{\delta\phi_1^2}} \big |_{\phi_1=\phi_2=0}&=&
\f{\delta^2\Gamma}{\delta\phi_2^2} \big |_{ \phi_1=\phi_2=0} \non
\f{\delta^4\Gamma}{\delta\phi_1^4} \big |_{ \phi_1=\phi_2=0}&=&
3\f{\delta^4\Gamma}{\delta\phi_1^2\delta\phi_2^2} \big |_{ 
\phi_1=\phi_2=0}
\eeqa

and in the broken phase which is defined by shifting the fields,
$\phi_1=\sigma+v$ and $\phi_2=\pi$,

\beqal{wi2}
v\f{\delta^2\Gamma}{\delta\pi^2} \big |_{
\sigma=\pi=0}=\f{\delta\Gamma}{\delta\sigma} \big |_{ \sigma=\pi=0}
\eeqa

where the $\phi_1$ field has been shifted by a constant amount $v$ which
fixes the vacuum for the broken phase.

The set of Ward identities, equation (\ref{wi1}), can be verified to one
loop \cite{my}, showing that
the quantum theory is symmetric. However care must be taken in defining
the broken phase so that equation (\ref{wi2}) holds.

If one naively shifts to a translationally invariant vacuum, the Ward
identity, equation (\ref{wi2}) only holds when the order of the
continuum and the IR limits is
such that the IR
divergences on the RHS of (\ref{wi2}) are transmuted to UV divergences
i.e. the
vacuum is defined with $p=0$ before all loop calculations. It is
important
to note here that shifting fields  by a constant amount $v$ at the
tree-level itself causes the $\sigma$-tadpole amplitude to scale
with respect to $\Lambda$ in the same way as the
second case in section 2. Where as the  $\Gamma^{\pi\pi}$ amplitude has
the same $\Lambda$ dependence when the external momenta are put to zero
before all loop computations. This makes the broken phase
nonrenormalizable with the same number of counterterms as the symmetric
phase. This point will be elaborated at the end of this section.  
However in order to define a UV
renormalizable theory for all values of $p$ we must work with an infrared
cutoff and remove the cutoff only after all loop computations as the 
former case in section 2. To do so one has to shift the fields
by $v$ which is not a constant. $v$ may be set to a constant after all
loop calculations. This would mean that the singular field configuration,
defined as the vacuum, is approached as a limit. One can also leave $v$ as
a nonconstant leading to a translationally noninvariant vacuum. 
This later case was studied
in \cite{phase} where phase transitions when a finite number of momentum
modes condense
is studied. 
In the present paper a nonconstant $v$ acts as an infrared regulator and
help us to
avoid proplems as noted earlier.

The Ward identity in the case where $v$ is not a constant will now be
written as,

\beqal{wi3}
\int\f{d^4p}{(2\pi)^4} 
v(-p)\f{\delta^2\Gamma}{\delta\pi(p_1)\delta\pi(p)} \big |_{
\sigma=\pi=0}=\f{\delta\Gamma}{\delta\sigma(p_1)} \big |_{ \sigma=\pi=0}
\eeqa 

Explicit computations in this case shows that,

\beqa
\Gamma^{\sigma}=v_0\mu^2-\lambda v_0^3 &+& v_0
[-3\lambda(\Lambda^2+\mu^2ln(\f{\Lambda^2}{-\mu^2}))+IR+F]+\non
&+&v_0^3[\f{5}{2}ln(\f{\Lambda^2}{-\mu^2})+IR+F]
\eeqa

\beqa
\Gamma^{\pi\pi}=-p^2+\mu^2-\lambda
v_0^2 &+& [-3\lambda(\Lambda^2+\mu^2ln(\f{\Lambda^2}{-\mu^2}))+IR+F]+\non
&+&v_0^2[\f{5}{2}ln(\f{\Lambda^2}{-\mu^2})+IR+F]
\eeqa

Where IR are the infrared divergent terms and F are the finite terms.
It can be seen that the UV divergence structure of the two amplitudes are
exactly as the Ward identity (\ref{wi2}). The IR divergences appear when
we go to a
translationally invariant vacuum i.e. $v$ is set to a constant $v_0$ 
after all loop calculations.

\subsection{Renormalizability to all orders : Review of The commutative
case}

We now consider the commutative globally symmetric $O(N)$ scalar theory in
its symmetric
phase and review its renormalizability to all orders following
Polchinski \cite{pol}. We
discuss here
the set up of the RG equations and merely state the results, which will 
be necessary for the later parts of this section. The reader may refer to
\cite{pol} for more elaborate details and proofs. 

\beqa
S(\phi)=\int
\f{d^4p}{(2\pi)^4}[-\f{1}{2}\phi^\alpha(p)\phi^\alpha(-p)(p^2-\mu^2)K^{-1} 
(\f{p^2}{\Lambda^2})] + L_{int}(\phi) \non
L_{int}(\phi)=\int
d^4x[-\f{1}{2}\rho_1^{0}(\phi^\alpha(x))^2-\f{1}{2}\rho_2^{0}
(\partial_\mu\phi^\alpha(x))^2-
\f{1}{4}\rho_3^{0}(\phi(x)^\alpha\phi(x)^\alpha)^2]
\eeqa

where, $K(\f{p^2}{\Lambda^2})$ has a value of 1 for $p^2<\Lambda^2$
and vanishes rapidly at infinity. $\rho_a^{0}$ are the bare couplings
defined at an UV cutoff scale $\Lambda_0$. 

The generating functional for the theory with the cutoff, $\Lambda$ may be
written as

\beqa
Z[J,\Lambda]=\int {\cal D}\phi exp[\int
\f{d^4p}{(2\pi)^4}[&-&\f{1}{2}\phi^\alpha(p)\phi^\alpha(-p)(p^2-\mu^2)K^{-1}
(\f{p^2}{\Lambda^2})\non &+& J^{\alpha}(p)\phi^{\alpha}(-p)]  
+ L_{int}(\phi)]
\eeqa

\beqal{dz}
\Lambda\f{dZ[J,\Lambda]}{d\Lambda}=\int {\cal D}\phi [\int
\f{d^4p}{(2\pi)^4}[&-&\f{1}{2}\phi^\alpha(p)\phi^\alpha(-p)(p^2-\mu^2) 
\Lambda\f{\partial K^{-1}}{\partial\Lambda}
(\f{p^2}{\Lambda^2})\non &+& J^{\alpha}(p)\phi^{\alpha}(-p)]
+ \Lambda\f{\partial L_{int}(\phi)}{\partial\Lambda}]exp(S(\phi))
\eeqa

The RHS of equation (\ref{dz}) vanishes if $L$ varies as,

\beqa
\Lambda \f{\partial L}{\partial\Lambda}=&-&\f{1}{2}\int d^4p
(2\pi)^4(p^2-\mu^2)^{-1}\Lambda \f{\partial K}{\partial\Lambda}
\non&[&\f{\partial L}{\partial\phi^{\alpha}(-p)} \f{\partial L} 
{\partial\phi^{\alpha}(p)}+\f{\partial^{2}L} 
{\partial\phi^{\alpha}(-p)\partial\phi^{\alpha}(p)}]
\eeqa

$L$ can now be expanded in terms of its Fourier modes. The global $O(N)$
symmetry
of the quantum theory implies that we can arrange the expansion as
follows.

\beqal{lexp}
L=\sum_{n=1}^{\infty}\f{1}{2n!}\int\prod_{i=1}^{n}\prod_{j=n+1}^{2n}d^4p_i
d^4p_j \phi^{\alpha}(p_i)\phi^{\alpha}(p_j)L_{2n}(p_1...p_{2n},\Lambda)
\delta^4(\sum_{i,j}p_i+p_j) 
\eeqa

Define the relevant operators as,
\beqa
\rho_1(\Lambda)=-L_2(p,-p,\Lambda)\big |_{p^2=p_0^2}\\
\rho_2(\Lambda)=-\f{\partial^2}{\partial p^2}L_2(p,-p,\Lambda)\big
|_{p^2=p_0^2}\\
\rho_3(\Lambda)=-L_4(p_1,p_2,p_3,p_4,\Lambda)\big |_{p={\bar p}}
\eeqa

Now, construct $V(\Lambda)$ such that,
\beqa
V(\Lambda)&=&\Lambda_0 \f{\partial
L(\Lambda)}{\partial\Lambda_0}-\sum_{a,b}
\f{\partial L(\Lambda)}{\partial\rho_{a}^{0}}\f
{\partial\rho_{a}^{0}}{\partial\rho_{b}(\Lambda)}
\Lambda_0\f{\partial\rho_{b}(\Lambda)}{\partial\Lambda_0}\non
\eeqa

\noindent
where, $a,b$ runs from 1 to 3. To prove that the theory is renormalizable
or in other words to show that, for the low energy theory to be finite one
has to tune a {\it finite} number of parameters,  $V(\Lambda)$ must be
shown to vanish at the low energy limit, $\Lambda/\Lambda_0\rightarrow 0$.

Expanding the following quantities similar to equation (\ref{lexp}),
\beqa
\f{\partial L(\Lambda)}{\partial\rho_{b}(\Lambda)}&=&\sum_{a}
\f{\partial L(\Lambda)}{\partial\rho_{a}^{0}}\f
{\partial\rho_{a}^{0}}{\partial\rho_{b}(\Lambda)}\non
&=&\sum_{n=1}^{\infty}\f{\Lambda^{4-2n-2\delta_{b1}}}{2n!} 
\int\f{\prod_{i=1}^{n}\prod_{j=n+1}^{2n}d^4p_i
d^4p_j}{(2\pi)^{8n-4}} 
\phi^{\alpha}(p_i)\phi^{\alpha}(p_j)\times\non 
&\times& B_{b,2n}(p_1...p_{2n},\Lambda)
\delta^4(\sum_{i,j}p_i+p_j)
\eeqa

\beqa
V(\Lambda)&=&\sum_{n=1}^{\infty}\f{\Lambda^{4-2n}}{2n!}
\int\f{\prod_{i=1}^{n}\prod_{j=n+1}^{2n}d^4p_i   
d^4p_j}{(2\pi)^{8n-4}}
\phi^{\alpha}(p_i)\phi^{\alpha}(p_j)\times
\non &\times& V_{2n}(p_1...p_{2n},\Lambda)
 \delta^4(\sum_{i,j}p_i+p_j)
\eeqa

Now defining,

\beqa
Q(p,\Lambda,\mu^2)=\f{1}{(p^2-\mu^2)}\Lambda^3\f{\partial
K(\f{p^2}{\Lambda^2})}{\partial\Lambda}
\eeqa

one arrives at the RG equations for $L,B,V$, shown in the appendix,
equations (\ref{rgl1}),(\ref{rgb1}),(\ref{rgv1}) where,

\beqa
\parallel f(p_1,...,p_{2n},\Lambda)\parallel=\max_{p_i^2<c\Lambda^2}|
f(p_1,...,p_{2n},\Lambda)|
\eeqa
\noindent

so that,

\beqal{qcomm}
\parallel \int \f{d^4p}{(2\pi)^4}Q(p,\Lambda)
L_{2n}(p_1,...,p_{2n-2},p,-p,\Lambda)\parallel  < C\Lambda^4
\parallel L_{2n}(\Lambda)\parallel
\eeqa

where, $c$ and $C$ are constants independent of $\Lambda$.
Now define the following initial conditions for the couplings,
\beqal{bccom}
\rho_1(\Lambda_R,\Lambda_0,\rho^0)&=&0\non
\rho_2(\Lambda_R,\Lambda_0,\rho^0)&=&0\non
\rho_3(\Lambda_R,\Lambda_0,\rho^0)&=&6\lambda_R
\eeqa

\noindent
Note that because of the $O(N)$ symmetry of the quantum theory it is
sufficient to set boundary conditions as above.
The irrelevant couplings are set to vanish at $\Lambda_0$.
Perturbative renormalizability now means that order by order in
$\lambda_R$ the following limit exists,

\beqa
\lim_{\Lambda_0\rightarrow \infty} {\bar
L}(\phi,\Lambda_R,\lambda_R,\Lambda_0)={\bar 
L}(\phi,\Lambda_R,\lambda_R,\infty)
\eeqa
where,

\beqa
{\bar L}(\phi,\Lambda_R,\lambda_R,\Lambda_0)= 
L(\phi,\Lambda_R,\Lambda_0,\rho^0(\Lambda_R,\lambda_R,\Lambda_0))
\eeqa

and specifically to $rth$ order in $\lambda_R$,

\beqa
\parallel {\bar L}_{2n}^{(r)}(\Lambda_R,\Lambda_0)-{\bar
L}_{2n}^{(r)}(\Lambda_R &,&\infty)\parallel\non
&\leq&\Lambda_{R}^{4-2n}(\f{\Lambda_R}{\Lambda_0})^2
P^{2r-n}ln(\f{\Lambda_R}{\Lambda_0}), r+1-n\geq 0\non
&=& 0 , r+1-n <  0
\eeqa

\noindent
where, $P^{2r-n}$ is a polynomial of degree $2r-n$.
We do not include the proof here, for which the reader may refer to the
original paper \cite{pol}, but merely state that it follows by induction
from the
following assertions,

\noindent
{\it (i)} At order $r$ in $\Lambda_R$,
\beqal{amax}
\parallel\partial_{i_1,j_1}^{\mu_1}...\partial_{i_p,j_p}^{\mu_p}A_{2n}^{(r)}
(p_i,...,p_{2n},\Lambda)\parallel &\leq&
\Lambda^{-p}P^{2r-n}ln(\f{\Lambda_0}{\Lambda_R}), r+1-n\geq 0\non
&=& 0, r+1-n<0
\eeqa
\noindent
where, $A_{2n}=\Lambda^{2n-4}L_{2n}$ and,
\beqa
\left( \f{\partial}{\partial p_i^{\mu}}-\f{\partial}{\partial p_j^{\mu}}
\right )A_{2n}=\partial_{i,j}^{\mu}A_{2n}
\eeqa
\noindent
{\it (ii)} At order $r$ in $\Lambda_R$,
\beqal{bmax}
\parallel\partial_{i_1,j_1}^{\mu_1}...\partial_{i_p,j_p}^{\mu_p}B_{b,2n}^{(r)}
(p_i,...,p_{2n},\Lambda)\parallel &\leq&
\Lambda^{-p}P^{2r-n+1+\delta_{b3}}ln(\f{\Lambda_0}{\Lambda_R}), r+2-n\geq 
 0\non
&=& 0, r+2-n<0
\eeqa

\noindent
{\it (iii)} At order $r$ in $\Lambda_R$,
\beqal{vmax}
\parallel\partial_{i_1,j_1}^{\mu_1}...\partial_{i_p,j_p}^{\mu_p}V_{2n}^{(r)}
(p_i,...,p_{2n},\Lambda)\parallel &\leq&
\Lambda^{-p}(\f{\Lambda}{\Lambda_0})^2P^{2r-n}ln(\f{\Lambda_0}{\Lambda_R}),
r+1-n\geq 0\non
&=& 0, r+1-n<0
\eeqa

\noindent
We end this part of the section by writing down the tree-level and the one
loop forms for the relevant and irrelevant parts of $L$. 
We use (\ref{rgl1}) to obtain the perturbative
expansion for $L_{2n}$ order by order in $\lambda_R$.

Expanding the two and four point parts of $L$,
\beqa
L_2(p,-p,\Lambda)&=&L_2(p,-p,\Lambda)\big
|_{p^2=p_0^2}+(p^2-p_0^2)\f{\partial}{\partial p^2}L_2(p,-p,\Lambda)\big
|_{p^2=p_0^2}+\Delta L_2 \non
&=&-\rho_1(\Lambda)-(p^2-p_0^2)\rho_2(\Lambda)+\Delta L_2\\
L_4(p_i,\Lambda)&=&L_4(p_i,\Lambda)\big |_{p_i={\bar p}_i}+\Delta L_4\non
&=&-\rho_3(\Lambda)+\Delta L_4
\eeqa

Solutions for equation (\ref{rgl1}) with boundary conditions
(\ref{bccom}) for zero and one loop
gives,

\beqa
\rho_{1}^{(0)}=0,\rho_{2}^{(0)}=0,\rho_{3}^{(0)}=6\lambda_R\\
\Delta L_2^{(0)}=0,\Delta L_4^{(0)}=0\\
\eeqa

\begin{figure}[htbp]
\begin{center}
\epsfig{file=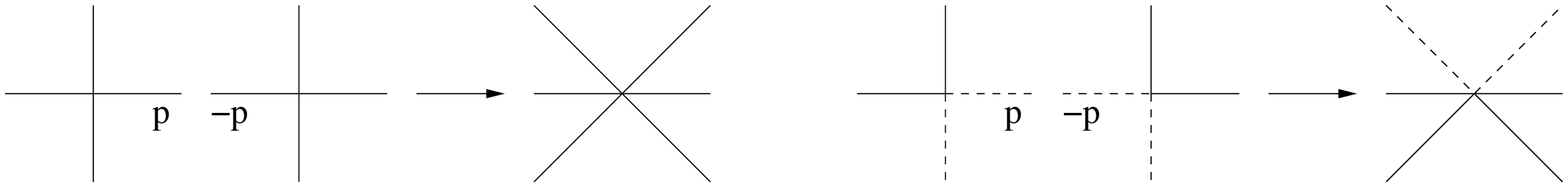, width=12 cm, angle=0}
\end{center}
\caption{$L_6$ vertex from $L_4$ vertices}
\end{figure}

The $L_6^{(0)}$ vertex is obtained from two $L_4^{(0)}$ vertices as shown
in Figure 4.
\beqa
L_6^{(0)}=\int_{\Lambda}^{\Lambda_0}\f{d\Lambda^{'}}{\Lambda^{'3}}
Q&(&\Lambda^{'},P)L_4^{(0)}(p_1,p_2,p_3,P,\Lambda^{'}) 
L_4^{(0)}(p_4,p_5,p_6,-P,\Lambda^{'})\non
&+&\f{1}{2}\left(\begin{array}{c}
6\\3\end{array}\right)-1 \mbox{ permutations for all same external fields 
or,}
\non &+& \f{1}{2}\left(\begin{array}{c}  
4\\2\end{array}\right) \mbox{ permutations for 2 different external
fields}
\eeqa

\noindent
where, $P=p_1+p_2+p_3$

Contracting two of the legs of the $L_4^{(0)}$ vertices we get the one
loop $L_2$ amplitude, Figure 5. The weights of $N$ coming from the
global
$O(N)$ symmetry are indicated in brackets in the figure. 

\begin{figure}[htbp]
\begin{center}
\epsfig{file=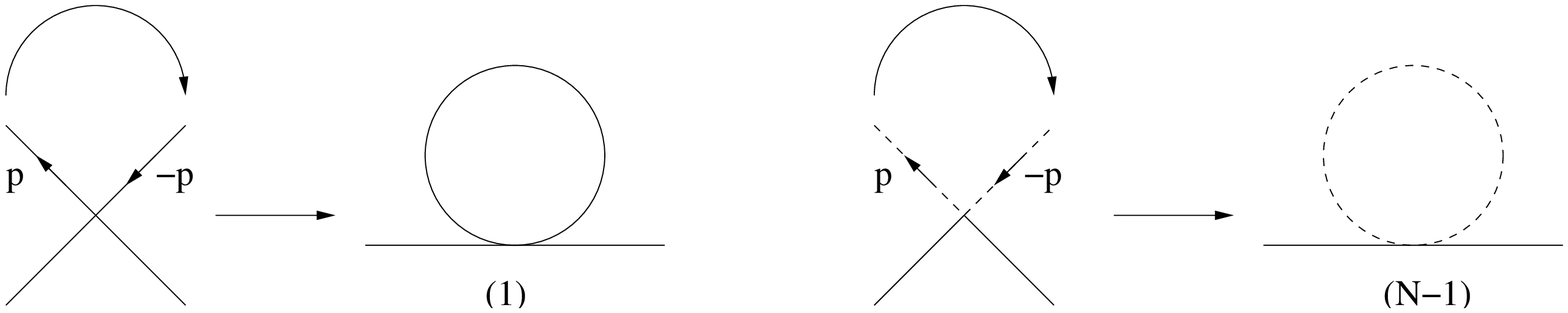, width=12 cm, angle=0}
\end{center}
\caption{One loop $L_2$ from $L_4$ vertex}
\end{figure}

\beqa
\rho_{1}^{(1)}&=&\f{1}{2}\int \f{d^4p} 
{(2\pi)^4}\int_{\Lambda_R}^{\Lambda}
\f{d\Lambda^{'}}{\Lambda^{'3}}Q(\Lambda^{'},p)L_4^{(0)}(p,-p,q,-q,\Lambda^{'})
\big |_{q^2=p_0^2}\non
&=&(N+2)\f{\lambda_R}{16\pi^2}\Lambda^2
\eeqa
\noindent
where we have only kept the $\Lambda$ dependent term.
\beqa
\rho_{2}^{(1)}&=&0, \Delta L_{2}^{(1)}=0\\
\eeqa

Contracting two legs of the $L_6^{(0)}$ vertex, Figure 6, we obtain the
one
loop $L_4$ amplitude. The $\Lambda$ dependent part of $\rho_{3}^{(1)}$
comes from this term only.

\begin{figure}[htbp]
\begin{center}
\epsfig{file=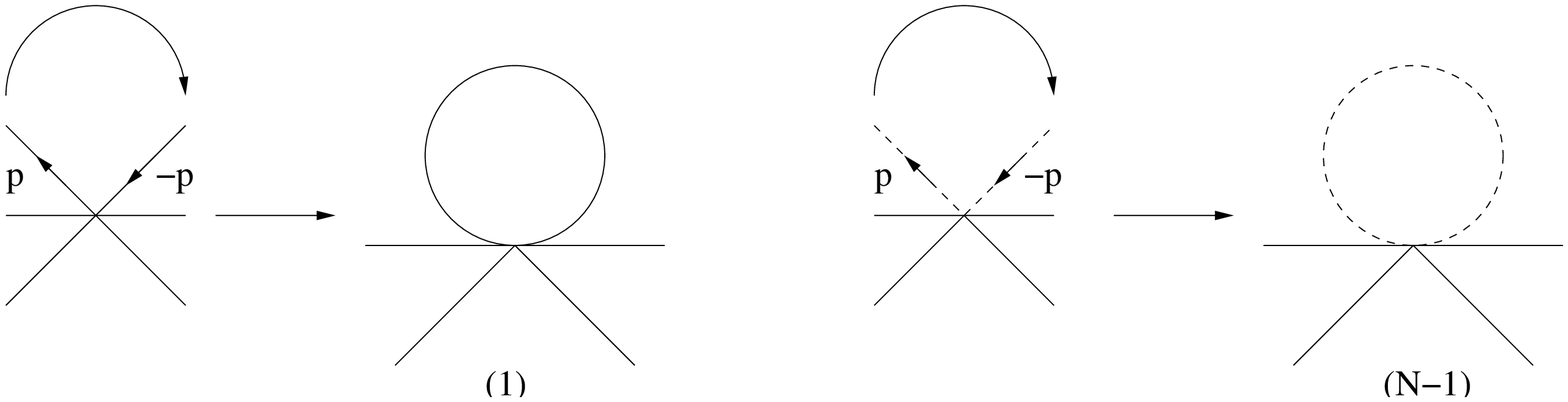, width=12 cm, angle=0}
\end{center}
\caption{One loop $L_4$ from $L_6$ vertex }
\end{figure}

\beqal{4ptc}
\rho_{3}^{(1)} &\sim& -\f{1}{2}\int
\f{d^4p}{(2\pi)^4}\int_{\Lambda_R}^{\Lambda}\f{d\Lambda^{'}}{\Lambda^{'3}}
Q(p,\Lambda^{'})L_6^{(0)}(p_1,p_2,p_3,p_4,p,-p,\Lambda^{'})
\big |_{p_i={\bar p}_i} 
\eeqa

In the large $\Lambda$ limit,
\beqa
\rho_{3}^{(1)}\sim -
(N+8)\f{3\lambda_R^2}{8\pi^2}ln(\f{\Lambda^2}{-\mu^2})
\eeqa

\beqa
\Delta L^{(1)}_4 \sim \f{f(\bar{p}_i)}{\Lambda}
\eeqa

\noindent
$f(\bar{p}_i)$ is a function of external momenta.
The relevant and the irrelevant operators scale at zero and one loop in the 
way asserted by equation (\ref{amax}) which is the primary step towards
proving the renormalizability
of the theory order by order in the loops by induction.


\subsection{The noncommutative case} 

We have seen that the $O(N)$ symmetry of the quantum theory enables
us to use the same set of RG equations for any $N$ as the $N=1$ case. 
We now discuss the renormalizability of the noncommutative theory to all
orders in its symmetric as well as in its broken phases. Apart from the 
UV cutoff $\Lambda$, we shall also introduce an IR cutoff, $\Lambda_{IR}$ 
as mentioned earlier.

First note that the components of $L$, $L_{2n}(p_1,...p_{2n},\Lambda)$
contain a phase factor which accounts for the noncommutavity of the
theory. This factor is of the form $e^{-\f{i}{2}\sum_{i<j}p_i\wedge p_j}$
with all possible permutations of the external momenta. Now consider
equation (\ref{qcomm}),

\beqal{qnoncom}
\parallel \int \f{d^4 p}{(2\pi)^4}
Q(p,\Lambda) L_{2n}&(&p_1,...p_{2n},p,-p,\Lambda)\parallel + \mbox{ all
possible permutations of $p_i$}\\ &=& \parallel
\int 
\f{d^4p}{(2\pi)^4} Q(p,\Lambda)e^{-\f{i}{2}\sum_{i<j}p_i\wedge p_j}
\parallel . \parallel
\tilde{L}_{2n}(p_1,...p_{2n-2},p,-p,\Lambda)\parallel
\non
&+&  \mbox{all 
possible permutations of $p_i$}
\non
\eeqa

where $\tilde{L}_{2n}(p_1,...p_{2n-2},p,-p,\Lambda)$ is the part
of
$L_{2n}$ not
containing the phase factor. Equation (\ref{qnoncom}) corresponds to the
one loop
vertex $L_{2n-2}$ evaluated from the $L_{2n}$ vertex.

Evaluating (\ref{qnoncom}) for one particular permutation of the external
momenta
gives,

\beqal{qnoncom1}
\int
\f{d^4p}{(2\pi)^4} Q(p,\Lambda)e^{-\f{i}{2}\sum_{i<j}p_i\wedge p_j}
e^{i(p\wedge \sum_{l}p_l)} \sim e^{-\f{i}{2}\sum_{i<j}p_i\wedge p_j}
 \f{\Lambda^4}{1+\Lambda^2 (\sum_{l}p_l) o (\sum_{l}p_l)}
\eeqa

The term on the RHS of equation (\ref{qnoncom1}) is the nonplanar
term. Depending upon
whether the two external momenta being contracted are consecutive to each
other or not we get the usual commutative planar term or nonplanar
terms respectively. See Figure 7.

\begin{figure}[htbp]
\begin{center}
\epsfig{file=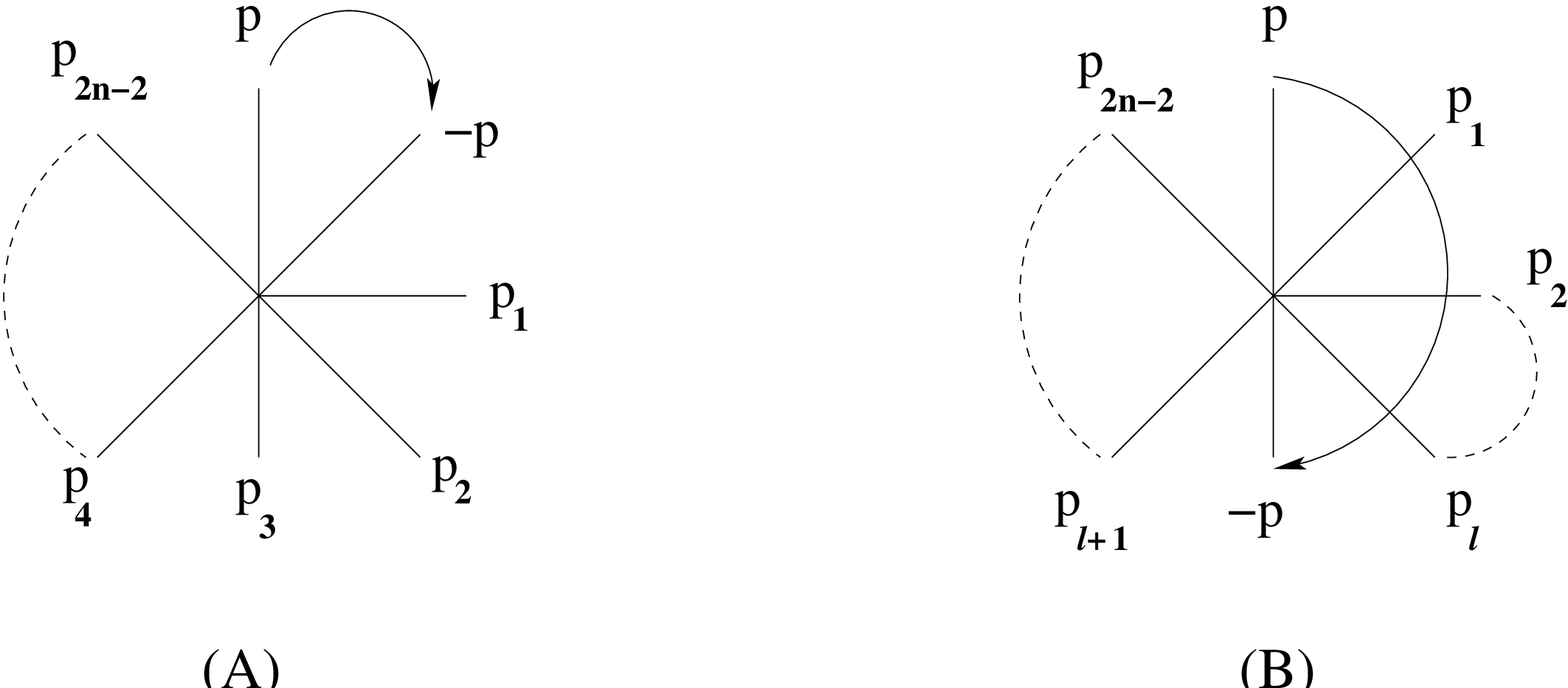, width=12 cm, angle=0}
\end{center}
\caption{Contractions leading to planar and nonplanar terms are shown in
(A) and (B) respectively}
\end{figure}

The exact weights of the planar and the nonplanar graphs would not be
necessary for our following discussions. We just label the weight of the
planar graph by $N_p$. For all permutations of the external
momenta, the scaling behavior of (\ref{qnoncom1}) with respect to
$\Lambda$ is,

\beqal{qnoncom2}
N_p \Lambda^4 + \f{\Lambda^4}{1+\Lambda^2 (\sum_{l}p_l) o (\sum_{l}p_l)}
 + ... \mbox {other nonplanar parts}
\eeqa

At this point it may be noted that, with every external momentum put to
zero the weight of the planar term changes due to the transmutation of the
nonplanar term into the planar term. 

We now turn to the question of UV
renormalizability of the theory by discussing the assertions as stated in
(\ref{amax}), (\ref{bmax}) and (\ref{vmax}). First, let us consider the
case where
all the
external momenta are
such that $\Lambda^2 pop >> 1$, $p$ being some combination of external
momenta. In this limit the nonplanar term in (\ref{qnoncom2}) scales as
$\Lambda^2$. The
scaling behavior of (\ref{qnoncom2}) is thus dominated by the planar term
for large
$\Lambda$. Therefore in this case the proof of assertion (i) follows
exactly as
the commutative case. Assuming that the assertion holds for $r=s-1$, 
from the RG equation (\ref{rgl1}), we have,

\beqal{rglp}
\parallel (\Lambda \f{\partial}{\partial
\Lambda}\partial^{\mu_1}_{i_1,j_1}...\partial^{\mu_p}_{i_p,j_p}
L_{2n}^{(s)}&(& p_1,...,p_{2n},\Lambda)\parallel \\ &\leq&
\Lambda^{4-2n-p}
[N_p + \f{1}{1+\Lambda^2 (\sum_{l}p_l) o (\sum_{l}p_l)}  
 + ...]P^{2s-n-1}ln(\f{\Lambda_0}{\Lambda_R}) \nonumber
\eeqa

For the case considered,  $\Lambda^2 pop >> 1$, we have,

\beqa
\f{1}{1+\Lambda^2 (\sum_{l}p_l) o (\sum_{l}p_l)} \sim
\f{1}{\Lambda^2 (\sum_{l}p_l) o (\sum_{l}p_l)}  
\eeqa

The two powers of $\Lambda$ in the denominator, decreases the overall
power of $\Lambda$ for $L_{2n}$ in the nonplanar terms so that the overall
scaling of $L_{2n}$ with respect to $\Lambda$ for large $\Lambda$ is
dominated by the planar term.

For $n\geq 3$ and  $n=2, p\geq 1$ ,the boundary values are set to zero at 
$\Lambda =\Lambda_0$. Therefore,

\beqal{lp}
\parallel \partial^{\mu_1}_{i_1,j_1}...\partial^{\mu_p}_{i_p,j_p}
L_{2n}^{(s)}&(& p_1,...,p_{2n},\Lambda)\parallel \\ &\leq&
\int_\Lambda^{\Lambda_0}\f{d\Lambda^{'}}{\Lambda^{'}}
(\Lambda^{'})^{4-2n-p}[N_p + \f{1}{\Lambda^{'2} (\sum_{l}p_l) o
(\sum_{l}p_l)}
 + ...]P^{2s-n-1}ln(\f{\Lambda_0}{\Lambda_R}) \non
&\leq& \Lambda^{4-2n-p}P^{2s-n-1}ln(\f{\Lambda_0}{\Lambda_R})
\nonumber
\eeqa

\noindent
In general $L^{s-1}_{2n+2}$ would have nonplanar terms as shown
above. When evaluating the  $L^{s}_{2n}$ vertex from  $L^{s-1}_{2n+2}$,
one has to integrate over all momenta thus including the small momentum
modes.
For these soft modes, $k$ one can expand the nonplanar term as,

\beqal{expan}
\f{\Lambda_{eff}(k)}{\Lambda^2}=\f{1}{1+\Lambda^2 kok}
=[1-\Lambda^2 kok + (\Lambda^2 kok)^2 ...]
\eeqa

\noindent
such that $\Lambda^2 kok \sim O(1)$. However in
this region of internal momenta, $k \sim O(1/\theta \Lambda)$, the
integral (\ref{qnoncom1}) without further exponential suppression is,

\beqal{}
e^{-\f{i}{2}\sum_{i<j}p_i\wedge p_j}\int^{\f{1}{\theta\Lambda}}_{0} \f
{d^4k}{(2\pi)^4}
Q(k,\Lambda)[1-\Lambda^2 kok +
(\Lambda^2 kok)^2 ...]  
\eeqa

The contribution from this momentum shell is suppressed by
$1/\theta\Lambda$. For larger values of
$k$ the integral is again suppressed by powers of $\Lambda$ in the
denominator.
Therefore even though the very small momentum modes in the internal lines
are included, it does not affect the cannonical scaling of the irrelevant
operators. We restrict ourselves to configurations of external momenta,
$p$ such that $\Lambda^2 pop >1$ because relevant operators scale with
different coefficients from the configurations with $\Lambda^2 pop <1$
and in the continuum limit IR divergences appear when the external
momenta are put to zero. It is thus clear that there is really no need for
an IR cutoff,
$\Lambda_{IR}$ in the loop integrals. The integrals are IR finite since we
are always working with a finite UV cutoff $\Lambda$. However restriction
to configurations with external momenta such that $\Lambda^2 pop >1$
implies the presence of an IR cutoff for the external momenta. Due to this
reason the IR cutoff for the internal loop momenta would be formally
assumed. The IR divergences in loop integrals however appear in
perturbation theory. This is illustrated by an example
(eqn. \ref{2loop}) at the end of this
proof of UV renormalizability.

For $n=2, p=0$, the boundary values are set at $p_i=\bar{p}_i$ and 
$\Lambda = \Lambda_R$. So from (\ref{rglp}), 

\beqal{rgl4pb}
\parallel \Lambda \f{\partial}{\partial
\Lambda}L^{(s)}_{4}(\bar{p}_i,\Lambda)\parallel &\leq&
P^{2s-3}ln(\f{\Lambda_0}{\Lambda_R})
\eeqa
which gives,

\beqal{l4pb}
\parallel L^{(s)}_{4}(\bar{p}_i,\Lambda)\parallel
 &\leq& L^{(s)}_{4}(\bar{p}_i,\Lambda_R) +
P^{2s-2}ln(\f{\Lambda_0}{\Lambda_R})\non
&\leq& P^{2s-2}ln(\f{\Lambda_0}{\Lambda_R})
\eeqa

where  $L^{(s)}_{4}(\bar{p}_i,\Lambda_R)$ is a constant independent of 
$\Lambda$. $L^{(s)}_{4}(p_i,\Lambda)$ can now be constructed for a general
momentum configuration from (\ref{l4pb}) using the Taylor's expansion,

\beqal{l4p}
 L^{(s)}_{4}(p_i,\Lambda)=L^{(s)}_{4}(\bar{p}_i,\Lambda) +
\sum^{3}_{i,j=1} p^{\mu}_{i} p^{\nu}_{j} \int_{0}^{1}d\lambda(1-\lambda)
\partial^{\mu}_{i,4}\partial^{\nu}_{j,4}
L^{(s)}_{4}(p^{'}_i,\Lambda)|_{p^{'}_i=\lambda p}
\eeqa

The terms on the RHS of (\ref{l4p}) being bounded by (\ref{lp}) and
(\ref{l4pb}),
$L^{(s)}_{4}(p_i,\Lambda)$ is also bounded. The bound for the components
of $L$ for $n=1$ and $p=0,2$ follows along the same line as above.
This proves the assertion (i).

Let us now see the  scaling behaviors of the Green's functions with
respect to $\Lambda$ for a
momentum configuration where some or all the external momenta are
zero. As noted earlier, with each external momentum put to zero, the
weights of the planar terms increases due to the transmutation of the
nonplanar terms to the planar terms. Therefore although the components of
$L$ scale with same power of $\Lambda$ as for the configuration with all
momenta nonzero, they scale with different
coefficients for the two configurations. Specifically
let us take the example of $L_4$ where $p_3=p_4=0$. The configuration of
fields for which
not all external momenta are nonzero is singular and so the scaling
behavior of functions for this configuration cannot be obtained from
(\ref{l4p}).
Instead one has to define the renormalization conditions at a point where
there are same number of zero external momenta so that, from
(\ref{rgl4pb}) we have

\beqa
\parallel L^{(s)}_{4}(\bar{p}_1,\bar{p}_2, 0,0,\Lambda)\parallel
 &\leq& L^{(s)}_{4}(\bar{p}_1,\bar{p}_2 ,0,0,\Lambda_R) +
P^{2s-2}ln(\f{\Lambda_0}{\Lambda_R})\non
&\leq& P^{2s-2}ln(\f{\Lambda_0}{\Lambda_R})
\eeqa

The constant $ L^{(s)}_{4}(\bar{p}_1,\bar{p}_2 ,0,0,\Lambda_R)$
differs from that of (\ref{l4pb}), but does not affect the scaling
behavior with
respect to $\Lambda$. 
$L_4(p_1,p_2,0,0,\Lambda)$ can now be obtained form
the Taylor's expansion about $(\bar{p}_1,\bar{p}_2 ,0,0)$.
However due to (\ref{qnoncom2}) with $p_3=0$, $p_4=0$    
$L^{(s)}_{4}(p_1,p_2 ,0,0,\Lambda)$ scales with a different   
weight of $ln(\f{\Lambda}{\Lambda_R})$ from that of (\ref{l4p}).
 
It is clear form these discussions that with the renormalization
conditions set at nonzero external momenta the Green's functions for
configurations of fields with some or all momenta zero cannot be
renormalized. This was the issue in the initial sections of this
paper. The bare couplings defined through the renormalization conditions
for the different configurations will have different $\Lambda_0$
dependences.
Again the different scaling of the greens funcions with respect
to $\Lambda$ for these configurations plays a crucial role in studying the
renormalizability of the broken phase of a spontaneously broken $O(N)$
theory. We shall discuss this at the end of this section.

The proof of assertions (ii) and (iii) follow along the same line as the
commutative case and the theory is renormalizable as long as we keep
away from the zero momentum limit. This concludes the proof of the
renormalizability of the noncommutative $O(N)$ symmetric theory.

We now give some computations of  zero and one loop contributions of the
relevant and
irrelevant parts of $L$ with the following boundary conditions.

\beqal{bcncom}
\rho_1(\Lambda_R,\Lambda_0,\rho^0)&=&0\non
\rho_2(\Lambda_R,\Lambda_0,\rho^0)&=&0\non
\rho_3(\Lambda_R,\Lambda_0,\rho^0)&=&2V({\bf \bar{p}})\lambda_R
\eeqa
\noindent
where $V({\bf \bar{p}})$ is given by (\ref{v}). The boundary values for
the
irrelevant couplings are set to zero at $\Lambda_0$ like the 
commutative case. As mentioned earlier the momentum integrals will now
also be regulated in the IR by $\Lambda_{IR}$. However in the expressions
for the one loop functions, to show the UV behavior, we drop all the
$\Lambda_{IR}$ dependent terms and retain only the UV cutoff dependent
pieces. 

\beqal{0loopnc}
\rho_{1}^{(0)}=0,\rho_{2}^{(0)}=0,\rho_{3}^{(0)}=2V({\bf
\bar{p}})\lambda_R\\
\Delta L_2^{(0)}=0,\Delta L_4^{(0)}=0
\eeqa

\beqal{2ptnc1}
\rho_1^{(1)}&=&-\lambda_R\int \f{d^4
p}{(2\pi)^4}\int_{\Lambda_R}^{\Lambda}
\f{d\Lambda^{'}}{\Lambda^{'3}}Q(\Lambda^{'},p)[N+1+cos(p\wedge q)]
\big |_{q^2=p_0^2}\non
&\sim&
-(N+1)\f{\lambda_R}{16\pi^2}[\Lambda^2+\mu^2ln(\f{\Lambda^2}{-\mu^2})]\non
&-&\f{\lambda_R}{16\pi^2}[\Lambda_{eff}^2(p_0)+\mu^2ln(\f{\Lambda_{eff}^2(p_0)}
{-\mu^2})]
\eeqa

\beqal{2ptnc2}
\rho_2^{(1)}&=&-\f{\partial}{\partial p^2}L_2^{(1)}(p,-p,\Lambda)
\big |_{p^2=p_0^2}\non
&\sim&-\f{\lambda_R}{16\pi^2}tr(\theta^2)[\Lambda_{eff}^4(p_0)
+\mu^2\Lambda_{eff}^2(p_0)]
\eeqa

\beqal{2ptnc3}
\Delta L_2^{(1)} &=& \f{\lambda_R}{16\pi^2}[\Lambda_{eff}^2(p)
+\mu^2ln(\f{\Lambda_{eff}^2(p)}{-\mu^2})]\big |_{\Lambda}^{\Lambda_0}\non
&-&\f{\lambda_R}{16\pi^2}[\Lambda_{eff}^2(p_0)
+\mu^2ln(\f{\Lambda_{eff}^2(p_0)}{-\mu^2})]
\big |_{\Lambda}^{\Lambda_0}\non
&+&\f{\lambda_R}{16\pi^2}(p^2-p_0^2)tr(\theta^2)
[\Lambda_{eff}^4(p_0)+\mu^2\Lambda_{eff}^2(p_0)]
\big |_{\Lambda}^{\Lambda_0}
\eeqa

\beqal{4ptnc}
\rho_3^{(1)}\sim (N+3)
V({\bf \bar{p}})\f{\lambda_R^2}{4{\pi}^2}ln(\frac{\Lambda^2}{-\mu^2}) +
G(\bar{p}_i)
\eeqa

$G(\bar{p}_i)$ contains the nonplanar terms which in the continuum
limit is only a function of the external momenta $\bar{p}_i$ and is
divergent at low momenta.

Equations (\ref{0loopnc}-\ref{4ptnc}) show the scaling of the relevant and
irrelevant operators with respect to $\Lambda$ at zero and one loop. As
long as the
external momenta are such that $\Lambda_0^2pop>1$, these parts scale as
asserted in (\ref{amax}). However we stress again that, ultimately one is
interested in the $\Lambda_0 \rightarrow \infty$ limit, so that these  
scaling behaviors persist as long as we keep away from the $p=0$ limit.

There are two points that may be noted here. For values of external
momenta, $p$ such that $\Lambda_0^2pop<<1$,

\noindent
{\it (i)} One can expand $\Lambda_{0eff}^2(p)$ in powers of
$\Lambda_0^2pop$, as in (\ref{expan}).
The irrelevant coupling (\ref{2ptnc3}) now is dependent on the UV cutoff
$\Lambda_0$, spoiling the usual Wilsonian picture as also noted in
\cite{wilson}.
Thus a renormalizable noncommutative quantum theory has to be defined with
an IR cutoff.\\
\noindent
{\it (ii)} The relevant couplings scale with different coefficients from
the noncommutative case where all external momenta are nonzero. 
They infact scale like the corresponding one loop
commutative theory couplings. In the limit $\Lambda^2 \bar{p}_i o
\bar{p}_j<<1$, the would be IR divergences terms in $G(\bar{p}_i)$,
equation (\ref{4ptnc}) are transmuted
into UV divergences and the noncommutative 4-point function scales with
respect to $ln(\Lambda^2)$ with the same coefficient as the commutative
4-point function (\ref{4ptc}).

Before discussing the renormalizability of the broken phase of the
spantaneously broken theory we now take a look at an example of the two
point function in perturbation theory and see the appearance of IR
divergences.\cite{min}

\begin{figure}[htbp]
\begin{center}
\epsfig{file=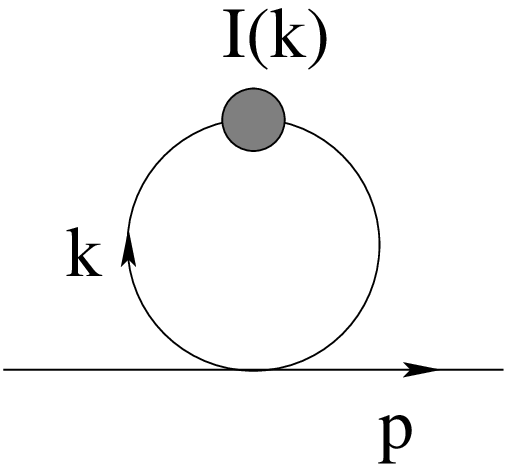, width=3 cm, angle=0}
\end{center}
\caption{A higher order two-point diagram in perturbation theory. $I(k)$
contains a number of tadpole insertions}
\end{figure}

The contribution from the diagram, shown in Figure 8 is given by,

\beqal{2loop}
\Gamma^{2} \sim \int d^4k \f{I(k)}{(k^2+m^2)^2} 
\eeqa

where the nonplanar part of $I(k)$ with $n$ tadpole insertions is given
by,

\beqa
I(k)=\f{1}{(\f{1}{\Lambda^2}+kok)^n} \f{1}{(k^2+m^2)^{n-1}}+ \mbox{less
singular terms}
\eeqa

The effect of high momenta in the tadpole insertions is encoded in
$\Lambda$. As all the high momenta modes are included, the singularities 
in $I(k)$ appears in the form of $1/(kok)^n$. The loop integral
\ref{2loop} is thus IR divergent. Note that this IR divergence in the loop
integrals never comes up in the RG approach, as the loop integrals are
always performed at a finite $\Lambda$.

We now turn to the issue of renormalizability of the broken phase in
the spontaneously broken theory. The broken phase of the theory is
defined by shifting the fields $\phi^{\alpha}$ such that,

\beqa
\phi^{\alpha}& \rightarrow&\sigma + v, \alpha=1\non
&\rightarrow&\pi^{\alpha},\alpha=2,N
\eeqa 

The two and the four point contributions to the effective lagrangian of
the broken phase is now given by,
\beqa
\f{1}{2}\int d^4p\phi^{\alpha}(p)\phi^{\alpha}(-p)L_2(p,&-&p,\Lambda)
\rightarrow \non \f{1}{2}\int d^4p[\sigma (p) \sigma (-p) &+& 2\sigma
(p) v(-p)+\pi^{\alpha}(p)\pi^{\alpha}(-p)]L_2(p,-p,\Lambda)
\eeqa

\beqa
\nonumber
\f{1}{4!}\int d^4p_1 d^4p_2 d^4p_3 d^4p_4 
[\phi^{\alpha}(p_1)\phi^{\alpha}(p_2)]
[\phi^{\alpha}(p_3)\phi^{\alpha}(p_4)&]&L_4(p_1,p_2,p_3,p_4,\Lambda)
\times\non &\times& \delta^4(\sum p_i)\rightarrow
\nonumber
\eeqa
\beqa
\f{1}{4!}\int d^4p_1 d^4p_2 d^4p_3 d^4p_4
&[&\sigma (p_1)v(p_2)v(p_3)v(p_4)+ permutations
+\pi^{\alpha}(p_1)\pi^{\alpha}(p_2)v(p_3)v(p_4)\non
&+&(p_1\rightarrow p_3,p_2\rightarrow p_4)+... ]
L_4(p_1,p_2,p_3,p_4,\Lambda)\delta^4(\sum p_i)
\eeqa

We have shown only these two contributions because these are the only
divergent functions.
The Ward identity (\ref{wi2}), which is the consequence of the $O(N)$
symmetry
of the symmetric phase tells that $v$ times the two point $\pi$-$\pi$
amplitude equals the $\sigma$-tadpole amplitude. This is manifest from the
above two expansions. However, from the foregoing analysis we have seen 
that the relevant couplings scale with different coefficients whether or
not the external momentum is such that $\Lambda_0^2pop<<1$. 
In the $\Lambda_0 \rightarrow \infty$ limit this inequality is satisfied
for $p\rightarrow 0$. The renormalization conditions at this limit defines
different $\Lambda_0$ dependences of the bare couplings from the
renormalization conditions at $p\ne 0$.

In particular, the contributions to the $\pi$-$\pi$ amplitude are from,

\beqal{pi}
L_2\pi^{\alpha}(p)\pi^{\alpha}(-p) +
L_4\pi^{\alpha}(p_1)\pi^{\alpha}(p_2)v(p_3)v(p_4) + (p_1\rightarrow
p_3,p_2\rightarrow p_4)
\eeqa
\noindent
and for the $\sigma$-tadpole amplitude from,
\beqal{sigma}
2L_2\sigma(p) v(-p)  +L_4\sigma(p_1) v(p_2)v(p_3)v(p_4) +
\mbox{permutations}
\eeqa 

We have seen that the behavior of Green's functions with respect to
$\Lambda$ changes once we put one or more external momenta to zero. The
same is true for the expressions (\ref{pi}) and (\ref{sigma}). Once $v$ is
set to a constant at the tree level the momenta associated with $v$
are set to zero. It is clear that in this case the $L_2$ and $L_4$
functions would have different $\Lambda$ dependences from the case where
$v$ is not a constant, for reasons outlined earlier. Now, since an unequal
number of $v$'s multiply (\ref{pi}) and (\ref{sigma}), the weights of the
UV divergent, $\Lambda$ dependent terms obtained due to transmutation of
the nonplanar terms to the planar terms by setting $v$ to a constant, are
different in (\ref{pi}) and (\ref{sigma}). This leads to a different
$\Lambda$ dependence in the $\sigma$-tadpole amplitude and the
$\pi$-$\pi$ amplitude. Consequently these two amplitudes cannot be
renormalized by the same counterterm, thus violating the Ward irentity
(\ref{wi2}). However the origin of the problem lies in the nonplanarity of
the diagrams induced by external momenta in the loop diagrams, the
resolution lies in keeping $v$ as a
nonconstant background field, so that the components of (\ref{pi}) and
(\ref{sigma})
scale in the same way and that the Ward identity (\ref{wi2}) holds.   
The symmetric phase of the theory has already been proved to be
renormalizable  as long as we have an infrared regulator. Spontaneous
symmetry breaking does not affect the renormalizability and the broken
phase is also renormalizable as long as $v$ is kept as a nonconstant
background field.

\section{Conclusion}

We have studied the UV renormalizability of noncommutative field theories.
In our discussions we have investigated this issue in the context of the 
$\lambda\phi^4$ theory, the Gross-Neveu model and the globally $O(N)$
symmetric $\phi^4$ model in its symmetric as well as its spontaneously
broken phases. The renormalizability of the globally $O(N)$ symmetric
$\phi^4$ theory is proved to all orders for both the symmetric as well as
the broken phases with an IR cutoff. 

The zero momentum configuration for these theories is singular. Through
our discussions of one loop results followed by a general analysis in the
language of the Wilsonian renormalization group, the following general
features of noncommutative field theories evolved.

\noindent
(i) We have seen that with the renormalization conditions set at a
momentum configuration where all the external momenta are nonzero, a
general Green's function with some or all the external momenta zero,
cannot
be renormalized. With each external momentum set to zero, the weights of
the UV divergent planar graphs increases due to UV/IR mixing. This leads
to a scaling behavior of the greens functions with respect to the cutoff 
($\Lambda$), with a different weight than a nonzero momenta configuration.
This implies that for the relevant operators, the bare couplings will have 
different UV cutoff dependences for these different configurations. \\
\noindent
(ii) The different scaling of greens functions for these seperate
configurations with respect to $\Lambda$ has crucial implications on the
renormalizability of the spontaneously broken phase of the $O(N)$
symmetric theory. In general the renormalizability of the broken phase of
the theory is unaffected by spontaneous symmetry breaking. 
The underlying $O(N)$  symmetry makes the broken phase
renormalizable with the same number of counterterms as the symmetric
phase. However in the case of noncommutative theory the one loop results
indicate that this only happens when we break the symmetry by going to a
vacuum which is translationally noninvariant. This can easily be
understood in the language of the Wilsonian renormalization group.
One of the consequences of the global $O(N)$ symmetry is the broken phase 
Ward identity (\ref{wi2}). It can be seen from (\ref{pi}) and
(\ref{sigma}), keeping in mind
the
scaling behaviors in the foregoing discussions, that the $\sigma$-tadpole
amplitude and the $\pi$-$\pi$ amplitude would scale differently with
respect to $\Lambda$ when $v$ is set to a constant. However when the
constant $v$ configuration is approached as a limit of the nonconstant $v$
configuration, the Ward identity (\ref{wi2}) is still preserved. We have
proved this to all orders following the proof of the symmetric phase of
the  $O(N)$ symmetric theory.  
\\
\noindent
(iii) There are no IR divergences in the loop integrals of the
Renormalization group equations. This is as a consequence of always
working with a finite UV cutoff $\Lambda$. IR divergences only show up
when we approach a singular, zero external momentum configuration after
taking the continuum limit in the solutions for the RG equations. If one
keeps away from these singular field configurations,
for generic external momenta, $p$, the theory is free from infrared
divergences. It is because of this reason and the scaling behaviors of
Green's functions in the two different momentum domains as discussed in
(i), that an IR cutoff for the external momenta is necessary. IR
divergences appearing in equation (\ref{2loop}) are thus artifacts of
perturbation theory.
\noindent
(iv) Finally (\ref{2ptnc3}) indicates that the irrelevant operators are
badly behaved
at the zero momentum configuration.

In the light of all these remarks it may be concluded that the zero
momentum configuration in noncommutative seems to make sense only when it
is approached as a limit from a nonzero momentum configuration, or in
other words, the noncommutative theories are renormalizable as long as
one works with an infrared cutoff.\\

\noindent
{\bf Acknowledgements}\\

\noindent
I am indebted to B. Sathiapalan for numerous discussions and
suggestions on the manuscript. I would also like to thank N. D. Haridass and 
H. S. Sharatchandra for valuable discussions and carefully reading the 
manuscript.

\appendix
\section{RG equations for $L$, $B$ and $V$}

RG equation for $L$ :
\beqal{rgl1}
\Lambda\f{\partial}{\partial\Lambda}L_{2n}=-\sum_{l=1}^{n}[\f{Q(P,\Lambda)}
{\Lambda^2}L_{2l}(p_1,...,p_{2l-1},P,\Lambda &)&
L_{2n+2-2l}(p_{2l},...,p_{2n},-P,\Lambda) + \non
&+&\f{1}{2}\left(\begin{array}{c}
2n\\2l-1\end{array}\right)-1 \mbox{ permutations}]
\nonumber
\eeqa
\beqa
-\f{1}{2}\int \f{d^4p}{(2\pi)^4}\f{Q(p,\Lambda)}{\Lambda^2}
L_{2n+2}(p_1,...,p_{2n},p,-p,\Lambda)
\eeqa

where, $P=\sum_{i=1}^{2l-1}p_i$, such that,

\beqal{rgl2}
\parallel(\Lambda^3\f{\partial}{\partial\Lambda}L_{2n})\parallel
\leq\sum_{l=1}^{n}[\f{1}{2}D_0\parallel L_{2l}(\Lambda)\parallel &.&
\parallel L_{2n+2-2l}(\Lambda)\parallel]\non
&+&\f{1}{2}C\Lambda^4\parallel
L_{2n+2}(\Lambda)\parallel
\eeqa

\noindent
where, $D,C$ are constants independent of $\Lambda$.

RG equation for $B$ :

\beqal{rgb1}
(\Lambda\f{\partial}{\partial\Lambda}&+&4-2n-2\delta_{b1})
B_{b,2n}(p_1,...,p_{2n})=\non 
&-& \sum_{l=1}^{n}[Q(P,\Lambda)A_{2l}(p_1,...,P,\Lambda)
B_{b,2n+2-2l}(p_{2l},...,p_{2n},\Lambda)+\left(\begin{array}{c}
2n\\2l-1\end{array}\right)-1 \mbox{ permutations}]\non
&-&\f{1}{2}\int\f{d^4p}{(2\pi\Lambda)^4}
B_{b,2n+2}(p_1,...,p_{2n},p,-p,\Lambda)Q(p,\Lambda)\non
&+& B_{1,2n}(p_1,...,p_{2n},\Lambda)\f{1}{2}\int \f{d^4q}{(2\pi\Lambda)^4}
B_{b,4}(p,-p,q,-q)Q(q,\Lambda)\big |_{p^2=p_0^2}\non
&+& B_{2,2n}(p_1,...,p_{2n},\Lambda)\f{\Lambda^2}{2}\int
\f{d^4q}{(2\pi\Lambda)^4}\f{\partial}{\partial p^2}
B_{b,4}(p,-p,q,-q)Q(q,\Lambda)\big |_{p^2=p_0^2}\non
&+& B_{3,2n}(p_1,...,p_{2n},\Lambda)\f{1}{2}\int
\f{d^4q}{(2\pi\Lambda)^4}
B_{b,6}(p_i,q,-q)Q(q,\Lambda)\big |_{p_i={\bar p}_i}
\eeqa

\beqal{rgb2}
\parallel(\Lambda\f{\partial}{\partial\Lambda}&+&4-2n-2\delta_{b1})
B_{b,2n}(p_1,...,p_{2n})\parallel
\non
&\leq& \sum_{l=1}^{n}[D_0\parallel A_{2l}(\Lambda)\parallel.
\parallel B_{b,2n+2-2l}(\Lambda)\parallel ]\non
&+&\f{1}{2}C\parallel B_{b,2n+2}(\Lambda)\parallel +\f{1}{2}C
\parallel B_{1,2n}(\Lambda)\parallel . \parallel B_{b,4}(\Lambda)
\parallel \non
&+&\f{1}{2}C\Lambda^2\parallel B_{2,2n}(\Lambda)\parallel . \parallel
\partial_{1,2}^{\mu}\partial_{1,2}^{\mu}B_{b,4}(\Lambda)
\parallel \non
&+&\f{1}{2}C
\parallel B_{3,2n}(\Lambda)\parallel . \parallel B_{b,6}(\Lambda)
\parallel 
\eeqa

RG equation for $V$ :

\beqal{rgv1}
(\Lambda\f{\partial}{\partial\Lambda}&+&4-2n)
V_{2n}(p_1,...,p_{2n})=\non
&-& \sum_{l=1}^{n}[Q(P,\Lambda)A_{2l}(p_1,...,P,\Lambda)
V_{2n+2-2l}(p_{2l},...,p_{2n},\Lambda)+\left(\begin{array}{c}
2n\\2l-1\end{array}\right)-1 \mbox{
permutations}]\non
&-&\f{1}{2}\int\f{d^4p}{(2\pi\Lambda)^4}
V_{2n+2}(p_1,...,p_{2n},p,-p,\Lambda)Q(p,\Lambda)\non
&+& B_{1,2n}(p_1,...,p_{2n},\Lambda)\f{1}{2}\int \f{d^4q}{(2\pi\Lambda)^4}
V_{4}(p,-p,q,-q)Q(q,\Lambda)\big |_{p^2=p_0^2}\non
&+& B_{2,2n}(p_1,...,p_{2n},\Lambda)\f{\Lambda^2}{2}\int
\f{d^4q}{(2\pi\Lambda)^4}\f{\partial}{\partial p^2}
V_{4}(p,-p,q,-q)Q(q,\Lambda)\big |_{p^2=p_0^2}\non
&+& B_{3,2n}(p_1,...,p_{2n},\Lambda)\f{1}{2}\int
\f{d^4q}{(2\pi\Lambda)^4}
V_{6}(p_i,q,-q)Q(q,\Lambda)\big |_{p_i={\bar p}_i}
\eeqa

\beqal{rgv2}
\parallel(\Lambda\f{\partial}{\partial\Lambda}&+&4-2n)
V_{2n}(p_1,...,p_{2n})\parallel
\non
&\leq& \sum_{l=1}^{n}[D_0\parallel A_{2l}(\Lambda)\parallel.
\parallel V_{2n+2-2l}(\Lambda)\parallel ]\non
&+&\f{1}{2}C\parallel V_{2n+2}(\Lambda)\parallel +\f{1}{2}C
\parallel B_{1,2n}(\Lambda)\parallel . \parallel V_{4}(\Lambda)
\parallel \non
&+&\f{1}{2}C\Lambda^2\parallel B_{2,2n}(\Lambda)\parallel . \parallel
\partial_{1,2}^{\mu}\partial_{1,2}^{\mu}V_{4}(\Lambda)
\parallel \non
&+&\f{1}{2}C
\parallel B_{3,2n}(\Lambda)\parallel . \parallel V_{6}(\Lambda)
\parallel
\eeqa

\end{document}